\documentstyle[twocolumn,aps,epsf]{revtex}

\begin{document}
%\preprint{HEP/123-qed}
\draft
\title{Stable,~Strongly~Attractive,~Two-State~Mixture~of~Lithium~Fermions~in~an~Optical~Trap}
\author{K. M. O'Hara, M. E. Gehm, S. R. Granade, S. Bali, and J. E. Thomas}
\address{Physics Department, Duke University,
Durham, North Carolina 27708-0305}
\date{\today}
\wideabs{\maketitle
\begin{abstract}
We use an all-optical trap to confine a strongly attractive
 two-state mixture of lithium fermions.
By measuring the  rate of evaporation from the trap, we determine the
effective
elastic scattering cross section  $4\pi a^{2}$ to show that the  magnitude
of the scattering length $|a|$  is very large, in agreement with predictions.
We show that the mixture is stable against  inelastic decay
  provided that a small  bias magnetic field  is applied. For this system,
  the s-wave interaction is widely tunable at low magnetic field, and can
be turned
  on and off  rapidly via  a  Raman $\pi$ pulse. Hence,
  this mixture is well suited for fundamental studies of  an interacting
Fermi gas.
 \end{abstract}

\pacs{PACS numbers: 32.80.Pj \\ Copyright 2000 by the American
Physical Society \\}}

Trapped, ultracold atomic vapors offer exciting new opportunities for
fundamental studies of an interacting Fermi gas in which the
temperature, density and interaction strength can be independently
controlled. Recently, a degenerate gas of
fermionic $^{40}{\rm K}$ has been produced by using a two-state
mixture  to enable s-wave scattering and
evaporation in a magnetic trap~\cite{Jin1}. By removing one species,
the properties of the
noninteracting degenerate gas were measured, demonstrating that
the momentum distribution and the total energy obey  Fermi-Dirac
statistics~\cite{Jin1}. However, the properties of
 interacting two-state fermionic vapors have not  been explored
 experimentally.

Theoretical treatments of an interacting Fermi gas have focused
extensively on  $^{6}{\rm
Li}$~\cite{BCS,Stringari,HighTC,Hydrodynamic,Parametric,Cooperpair,BruunBurnett,ConMatRev}.
Certain two-state $^{6}{\rm Li}$ mixtures  are  predicted to be
strongly attractive, i.e., they have  anomalously large and
negative scattering lengths~\cite{Elastic} arising from a
near-zero energy resonance in the triplet state~\cite{s-wave}. It
has been predicted
 that  these strongly attractive mixtures  can undergo a transition to
 a superfluid state at a relatively high transition
temperature~\cite{BCS,HighTC}.
   In addition, the two-state effective interaction
potential is widely tunable in a magnetic field, permitting
systematic studies of fundamental phenomena such as  collective
oscillations for both the normal and superfluid
phases~\cite{Stringari,Hydrodynamic,Parametric},  as well as new
tests of superconductivity theory~\cite{HighTC}.

Unfortunately, magnetically trappable mixtures in $^{6}{\rm Li}$
with large s-wave scattering  lengths are not stable, since
  there are correspondingly large spin-exchange and dipolar
 decay rates~\cite{BCS,Cooperpair,Elastic}. Hence, the methods
 employed to study degenerate $^{40}{\rm K}$ are not applicable.  For this
reason,
 we developed an ultrastable ${\rm CO}_{2}$
 laser trap to confine a stable mixture of
 the two lowest $^{6}{\rm Li}$ hyperfine states~\cite{O'Hara}.
 However, attaining a large and negative scattering length in this mixture
requires
 high magnetic fields  $B \geq  800$ G to exploit either a Feshbach
 resonance or the triplet scattering length~\cite{Cooperpair,Elastic}.

In this Letter, we show that there exists another stable hyperfine
state mixture in  $^{6}{\rm Li}$ which has the following unique
properties. First, we predict that the scattering length $a$ is
large,  negative, and widely tunable at {\em low} magnetic field
$B$. By monitoring the rate of evaporation from the ${\rm CO}_{2}$
laser trap at a fixed well depth, we measure
$|a|=540_{-100}^{+210}\,a_{0}$ at $B=8.3$ G. This result confirms
for the first time that very large scattering lengths exist in
$^{6}{\rm Li}$ mixtures. The predicted scattering length is
$-490\,a_{0}$ at $B=8.3$ G, consistent with our observations, and
is expected to increase to $-1615\,a_{0}$ as $B\rightarrow 0$.
Second, we find that this system is stable against spin exchange
collisions provided that $B\neq 0$. In addition, the dipolar decay
rate  is predicted to be very small~\cite{Verhaar}, consistent
with our observations.  Finally,  in the experiments, a   Raman
$\pi$ pulse is employed to abruptly create an interacting mixture
from a noninteracting one, a desirable feature  for studies of
many-body quantum dynamics~\cite{Zoller}.
\begin{figure}
\begin{center}\
\epsfysize=62mm \epsfbox{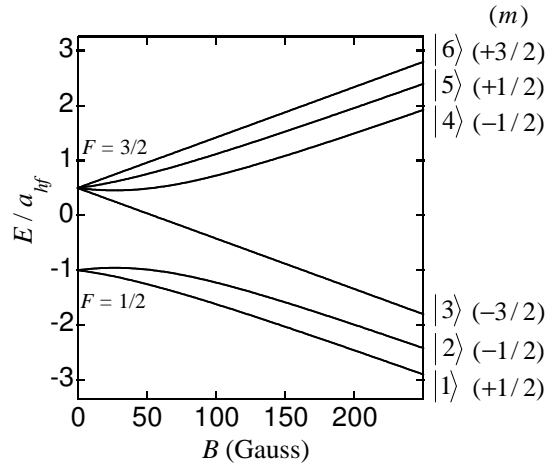}
\end{center}
\caption{$^{6}{\rm Li}$ hyperfine states,
labeled $|1\rangle$ to  $|6\rangle$ in order of
increasing energy in a magnetic field. The magnetic quantum
number of each state is denoted by $m$. The hyperfine constant
$a_{hf}=152.1$ MHz.}
\label{fig:levels}
\end{figure}
Fig.~\ref{fig:levels} shows the  hyperfine states for $^{6}{\rm Li}$
labelled $|1\rangle -|6\rangle$, in order of increasing energy in a
magnetic field. At low field, the states $|1\rangle$ and $|2\rangle$
correspond to the $|F=1/2,\,m\rangle$ states, while states
$|3\rangle$ through $|6\rangle$ correspond to
states $|F=3/2,\,m\rangle$. At nonzero
magnetic field, only the magnetic quantum number $m$ is conserved.
The subject of this paper is the $|3\rangle - |1\rangle$ mixture.
\begin{figure}
\begin{center}\
\epsfysize=62mm \epsfbox{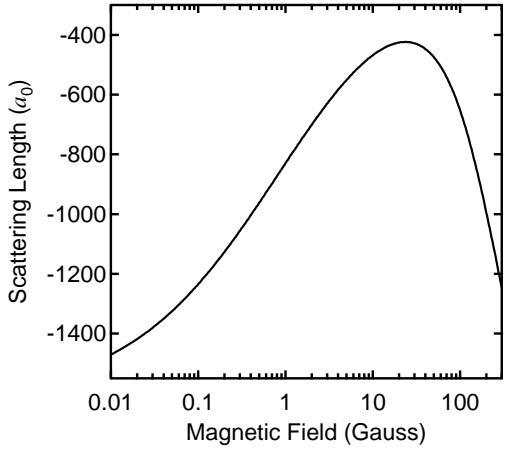}
\end{center}
\caption{Magnetic field dependence of the scattering length $a_{31}$
for a mixture of the  $|3\rangle$ and $|1\rangle$   hyperfine
states of $^{6}{\rm Li}$.}
\label{fig:a31}
\end{figure}
Fig.~\ref{fig:a31} shows  the scattering length $a_{31}$ for the
$|3\rangle - |1\rangle$ mixture as a function of magnetic bias
field $B$.  We estimate $a_{31}(B)$  by using the asymptotic
boundary condition (ABC) approximation~\cite{Elastic}. This
calculation incorporates  the singlet and triplet scattering
lengths~\cite{s-wave}, $a_{S}=45.5\pm 2.5\,a_{0}$ and
$a_{T}=-2160\pm 250\,a_{0}$, and a boundary radius which we take
to be $R=40\,a_{0}$~\cite{Elastic}. The scattering length varies
from $-1620\,a_{0}$ ($\simeq 3a_{T}/4$ as $B\rightarrow 0$) to
$-480\,a_{0}$ at $B=10$ G. The results of our approximate
calculation for $B=0$ to $B=200$ G are confirmed within 10\% by
van Abeelen and Verhaar~\cite{Verhaar} using a coupled channel
calculation which includes the uncertainties in the potentials. At
higher fields, near 800 G, we believe the scattering length
exhibits a Feshbach resonance (not shown). Above this resonance,
the scattering length approaches the triplet scattering length of
$-2160\,a_{0}$.

The $|3\rangle - |1\rangle$ mixture is stable against spin-exchange
collisions provided that a small bias magnetic field  is applied.
Spin-exchange inelastic collisions
conserve the  two-particle total magnetic quantum number  $M_{T}$,
where $M_{T}=-1$ for the $|\{3,1\}\rangle$ state.
Note that $\{,\}$ denotes the antisymmetric
 two-particle spin state, as  required for s-wave scattering which
 dominates at  low temperatures.
There are no lower-lying antisymmetric states with $M_{T}=-1$.
Hence, exothermic collisions are precluded. The only other states
with $M_{T}=-1$ are  $|\{4,2\}\rangle$ and $|\{5,3\}\rangle$.
Without an adequate bias magnetic field, transitions to these
states lead to population in level $|4\rangle$. Then, exothermic
$| \{3,4\}\rangle\rightarrow |\{3,2\}\rangle$ and $|\{4,1\}\rangle
\rightarrow |\{1,2\}\rangle$ collisions can occur. With an
adequate  bias magnetic field, the energy of states
$|\{4,2\}\rangle$ and  $|\{5,3\}\rangle$ can be  increased
relative to that of state $|\{3,1\}\rangle$  by more than the
maximum relative kinetic energy, i.e., twice the well depth during
evaporative cooling. By energy conservation, spin-exchange
transfer is then suppressed. In this case, the  inelastic rate is
limited to magnetic dipole-dipole (dipolar) interactions which
contain a rank 2 relative coordinate operator of even
parity~\cite{BCS}.  Since  parity is conserved and ${\rm
p-wave}\rightarrow {\rm p-wave}$ scattering is frozen out at low
temperature,  the dominant dipolar process is a small  ${\rm
s}\rightarrow {\rm d}$ rate in which
 $|\{3,1\}\rangle\rightarrow |\{1,2\}\rangle$~\cite{Verhaar}.

In the experiments, the ${\rm CO}_{2}$ laser trap is initially
loaded from a magneto-optical trap (MOT)\cite{O'Hara}. At the end
of the loading period, the MOT laser beams are tuned near
resonance and  the intensity is lowered to decrease the
temperature. Then, optical pumping is used to empty the $F=3/2$
state to produce a 50-50 mixture of the  $|1\rangle -|2\rangle$
states. These states are noninteracting at low magnetic field,
i.e, the scattering amplitude vanishes as a result of an
accidental cancellation~\cite{Elastic}. With a ${\rm CO}_{2}$
laser trap  depth of $330\,\mu$K, up to $4\times 10^{5}$ atoms are
confined in the lowest-lying hyperfine states  at an initial
temperature between 100 and 200 $\mu$K. A bias magnetic field of
8.3 G is applied to split the two-particle energy states by
$\simeq 16$ MHz. This is twice the maximum attainable energy at
our largest well depth of $400\,\mu{\rm K}=8$  MHz. After a delay
of 0.5 second relative to the loading phase, a pair of optical
fields is pulsed on to induce a Raman $\pi$ pulse. This pulse
transfers
 the population in state $|2\rangle$ to state $|3\rangle$ in two
 microseconds, initiating evaporative cooling in the resulting $|3\rangle
-|1\rangle$
mixture. The optical  fields are detuned from resonance with the D2
transition by
$\simeq 700$ MHz to suppress optical pumping. If the Raman pulse is
not applied, the trapped atoms remain in the noninteracting
$|1\rangle-|2\rangle$ mixture and exhibit a purely exponential decay
with a time constant $\simeq 300$ seconds.

An acousto-optic modulator (A/O) in front of the  ${\rm CO}_{2}$
laser controls the laser intensity, which is reduced to yield a
shallow trap depth of $100\,\mu$K. By using a shallow well, we
avoid the problem that the elastic cross section becomes
independent of the scattering length at high energy, as described
below.
 In addition, the shallow well greatly reduces the
number of loaded atoms and  makes  the sample
 optically thin, simplifying calibration of the number of trapped atoms.
 To determine the trap parameters,
the laser power is modulated and parametric resonances~\cite{Hansch} are
observed
at drive frequencies of $2\nu$ for three different trap oscillation
frequencies
$\nu$: At $100\,\mu$K well depth, $\nu_{x}=2.4$ kHz   $\nu_{y}=1.8$ kHz
and $\nu_{z}=100$ Hz, where the trap laser beam propagates along
${\bf z}$. Using the measured total power
as a constraint, we obtain the trap intensity $1/e^{2}$ radii,
 $w_{x}=50\,\mu$m and  $w_{y}=67\,\mu$m, and the axial intensity $1/e^{2}$
 length, $z_{f}\simeq 1.13$
 mm, where $z_{f}$ is consistent with the expected Rayleigh length within
15\%.

The  number  of atoms in the trap
$N(t)$ is estimated using a calibrated photomultiplier. The detection
 system monitors the  fluorescence induced  by  pulsed, retroreflected,
 $\sigma_{\pm}$  probe and repumper beams which are strongly saturating
   ($I/I_{sat}=26$  for the strongest transition).   To simplify  calibration,
   only the isotropic component of the fluorescence angular   distribution
  is measured: The  collecting  lens is placed at  the magic
angle~\cite{Magic} of
  $55^{\circ}$ ($P_{2}(\cos\theta )=0$)
  with respect to the propagation direction of the probe beams.
   The net efficiency of the detection system
    is determined using laser light of known power.
    The primary uncertainty in the calibration
arises from the excited state population fraction, which
we estimate lies between 1/4 and 1/2.
\begin{figure}
\begin{center}\
\epsfysize=62mm \epsfbox{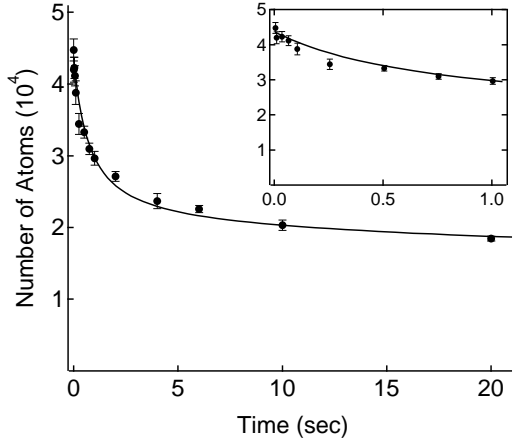}
\end{center}
\caption{Number of trapped atoms versus time for evaporation
of a $|3\rangle -|1\rangle$ mixture of $^{6}{\rm Li}$ at a fixed well depth
of $100\,\mu$K. The solid curve shows the s-wave Boltzmann equation
fit for a scattering length of $|a_{31}|=540\,a_{0}$. Inset: 0-1 second.}
\label{fig:number}
\end{figure}
Fig.~\ref{fig:number} shows the number of trapped atoms $N(t)$
 measured for the $|3\rangle -|1\rangle$ mixture at
 a well depth $U_{0}=100\,\mu$K and a bias field of 8.3 G
  as a function of time  between 5 ms and 20 seconds after  evaporation is
initiated.
   For times beyond 50 seconds (not shown), the
  evaporation stagnates, and   we observe an exponential decay of the
 cooled $|3\rangle -|1\rangle$
  mixture with a time constant of 370 seconds over a period   of a few
hundred seconds.
  The error bars are the standard deviation
of the mean of ten complete runs through the entire time sequence.

A model based on the s-wave Boltzmann equation~\cite{Walraven} is used
to predict $N(t)$ for comparison to the experiments.
This equation is modified
to include the density of states for a gaussian potential
well~\cite{O'Hara} and to include the energy dependence
of the elastic cross section. Assuming a short range potential
and a  symmetric (s-wave) spatial state,
the cross section takes the form
\begin{equation}
\sigma (k)=\frac{8\pi a_{31}^{2}}{1+k^{2}a_{31}^{2}},
\label{eq:cross}
\end{equation}
where  $\hbar k$ is the relative momentum. For $k|a_{31}|<<1$, the
cross section is maximized.
When $k|a_{31}|>>1$, the cross section approaches the unitarity limit
$8\pi /k^{2}$ which is independent of $a_{31}$.
Note that $k|a_{31}|=1$ corresponds to a relative kinetic
energy of $\epsilon =\hbar^{2}/(2\mu\,a_{31}^{2})$, where $\mu =M/2$ is
the reduced mass. For $|a_{31}|=500\,a_{0}$, $\epsilon =115\,\mu$K.

For a two-state mixture of fermions, the effective cross section
is reduced from that of Eq.~\ref{eq:cross} by a factor of 2
since  pairs of colliding  atoms  are in an antisymmetric hyperfine state with
a probability 1/2. This effective cross section is used in a Boltzmann
collision integral for each state $i=1,3$. A decay term
$-N_{i}(t)/\tau$ with $\tau =370$ sec
is added to account for the measured trap lifetime.
A detailed description of our  coupled Boltzmann equation
model will be published elsewhere.

The coupled s-wave Boltzmann equations for the two states
are numerically integrated to determine  $N(t)$ using
 the well parameters as fixed inputs.   From the calibrated
photomultiplier signal,
assuming that 1/3 of the atoms are in the  excited state,
 we obtain an initial total number $N_{0}=44,000$.
 For this case, the initial collision rate in Hz is   estimated to be
$1/(2\pi \tau_{c})\simeq N_{0}M\sigma_{0}\nu^{3}/(k_{B}T)$, where
$\nu^{3}=\nu_{x}\nu_{y}\nu_{z}$, $\sigma_{0}=8\pi a_{31}^{2}$, and
$M$ is the $^{6}{\rm Li}$ mass. Assuming $|a_{31}|=500\,a_{0}$,
$\tau_{c} =30$ ms. Hence, for $t>0.3$ seconds, when on average 10
collisions have occurred, the sample should be thermalized as
assumed in the theory.

 The best  fit to the data starting with  22,000 atoms in each
 state is shown as the solid curve in Fig.~\ref{fig:number}.
The $\chi^{2}$ per degree of freedom for this fit is  1.4 and is found
to be very sensitive to the initial temperature $T_{0}$ of the atoms
in  the optical trap.
>From the fit, we find  $T_{0}=46\mu$K, which is less than the well depth.
We believe that this low temperature is a consequence
of  the MOT gradient magnet, which  is turned off after the MOT laser
beams. The effective well depth of the optical trap
is therefore reduced until the gradient is fully off,
allowing hotter atoms to escape before the Raman pulse is applied to
create the $|3\rangle -|1\rangle$ mixture.
The fit is most sensitive to
data for $t>0.5$ second, where the thermal approximation is expected to
be valid. From the fit, we obtain the scattering length
 $|a_{31}|=540\pm 25\, a_{0}$, which is within 10\% of the predictions of
 Fig.~\ref{fig:a31}. The quoted error corresponds to
 a change of 1 in the total $\chi^{2}$.

We  determine the systematic errors in $a_{31}$ due to the
 uncertainties  in  the calibration and in the population imbalance
 as follows. The data is fit for  an
 initial number of atoms $N_{0}$ of  58,000 and  29,000,
 corresponding to an excited state fraction of
 1/4 and 1/2. This  yields $|a_{31}|=440\pm 20\,a_{0}$ and
$|a_{31}|=750\pm 42\,a_{0}$, respectively. Note that for the
larger scattering lengths, the cross section given by
Eq.~\ref{eq:cross} approaches the unitarity limit and the  error
increases. We assume that the initial population imbalance for
states $|3\rangle$ and $|1\rangle$ is comparable to that of states
$|2\rangle$ and $|1\rangle$ in the optically pumped MOT. To
estimate the latter population imbalance, we use state-selective
Raman $\pi$ pulses to excite $|2\rangle\rightarrow |3\rangle$ or
$|1\rangle\rightarrow |6\rangle$ transitions in the MOT.
Probe-induced  fluorescence signals from states $|3\rangle$ or
$|6\rangle$
 show that the initial $|1\rangle$ and $|2\rangle$ populations
 are  equal within 10\%. Note that  residual population in state
$|2\rangle$ is expected to be stable and weakly interacting, since we
estimate  $|a_{32}|<30\,a_{0}$ for  $0\leq B\leq 50$ G using
 the  ABC method, and $a_{12}\simeq  0$~\cite{Elastic}.
 Using the parameters for  the fit shown
in Fig.~\ref{fig:number}, but changing the initial mixture
from 50-50 to 60-40, we find a slight increase in the
fitted scattering length from $540\,a_{0}$ to $563\,a_{0}$.
Thus, the   uncertainty in the calibration of the number of atoms
produces the dominant uncertainty and $|a_{31}|=540^{+210}_{-100}\,a_{0}$.

To demonstrate that evaporative cooling is occuring, rather than just
trap loss, we have also measured the final temperature of the mixture
using release and recapture~\cite{Chu} from the ${\rm CO}_{2}$ laser trap.
We obtain $9.8\pm 1\,\mu$K, which is within 10 \% of the final
temperature of $8.7\,\mu$K predicted by the Boltzmann equation model.
An excellent fit to the data is obtained for the  final temperature,
which describes a thermal distribution.
However, the initial temperature is not so readily measured, as
it is nonthermal before evaporation is initiated, and is rapidly
changing during evaporation, unlike the final temperature, which
stagnates.

Good fits to the evaporation data are obtained  neglecting
inelastic collisions, suggesting that the dipolar rate for the
$|3\rangle -|1\rangle$ mixture is small, in contrast to the
scattering length. A limit on the dipolar loss  rate for the
$|3\rangle -|1\rangle$ mixture can be estimated from the $\tau
=370$ second  lifetime of  the mixture after evaporation
stagnates. For equal populations in both states, dipolar decay
results in an initial loss rate $\dot{n}=-Gn^{2}/4$, where $G$ is
the dipolar rate constant and $n$ is the total density. To obtain
a high density, the trap is loaded at a well depth of $330\,\mu$K
and the  temperature of the atoms is  reduced by evaporation to
$T\simeq 30\pm 1\,\mu$K.  The number of atoms remaining in each
state after evaporation is estimated to be $N=6.5\pm 2.2\times
10^{4}$, where the uncertainty is in the calibration. We cannot
rule out the possibility that one state is depleted on a long time
scale, since we do not directly measure the individual state
populations.  However, we believe that, after evaporation
stagnates in the deep well, a $|3\rangle-|1\rangle$ mixture
remains, since subsequent reduction of the well depth yields final
temperatures consistent with evaporative cooling.  Note that the
mixture ratio is not critical: An 80-20 mixture yields an initial
loss  rate $\dot{n}=-0.16\,Gn^{2}$, $\simeq 2/3$ that of a 50/50
mixture. For a fixed $330\,\mu$K trap depth, $\nu^{3}=2.6\pm
0.3\,{\rm kHz}^{3}$, and the phase space density for one state in
the harmonic approximation is then $\rho_{ph}=N/(k_{B}T/h\nu
)^{3}=7\times 10^{-4}$. This corresponds to a   maximum total
density of $n=2\,\rho_{ph}/\lambda_{B}^{3}=6.4\times 10^{11}/{\rm
cm}^{3}$, where $\lambda_{B}\equiv h/\sqrt{2\pi M\,k_{B}T}$. Since
the exponential decay time of the $|3\rangle -|1\rangle$ mixture
is similar to that obtained in the noninteracting $|1\rangle
-|2\rangle$ mixture, we assume the loss is dominated by background
gas collisions. Thus, we must have $Gn/4<<1/\tau$, which yields
$G<<2\times 10^{-14}{\rm cm}^{3}/{\rm sec}$.
 This result is consistent with the value
$G\simeq 2\times 10^{-15}\,{\rm cm}^{3}/{\rm sec}$ predicted
for the dipolar rate constant
at $30\,\mu$K  by van Abeelen and Verhaar~\cite{Verhaar}.

Future experiments will employ continuous evaporation
by slowly reducing the well depth~\cite{Adams}.
In this case, very large scattering lengths can be obtained
at low temperatures and small well
depths by using a reduced bias magnetic field $B$.  By adiabatically
recompressing the well, experiments can be carried out with the precooled
atoms
in a deep trap  to obtain high density as well. In such experiments, the final
low temperature  limits the number of atoms in the high energy
tail of the energy distribution,  exponentially
suppressing  spin-exchange collisions for $B\neq 0$.
For example, if  a total of  $3\times 10^{5}$ atoms were contained in our trap
at a well depth of $400\,\mu$K, the Fermi temperature $T_{F}=7\,\mu$K
and the Fermi density is $4\times 10^{13}/{\rm cm}^{3}$.
At a temperature of $T=0.1\, T_{F}=0.7\,\mu$K, a bias field of $B=0.16$ G
would split the two-particle hyperfine states by
$k_{B}T_{F}+12\,k_{B}T$, suppressing the spin exchange rate
 by  $\exp ( -12)$, and giving $a_{31}\simeq -1200\,a_{0}$.
 Alternatively, as shown in Fig.~\ref{fig:a31},
 large $a_{31}$ can be obtained at  moderate $B\simeq 300$ G.

In conclusion, we have observed that
an optically trapped $|3\rangle -|1\rangle$
mixture of $^{6}{\rm Li}$ atoms has a very large scattering length
at low magnetic field. This mixture is stable against spin-exchange
collisions provided that a small  bias magnetic field is applied.
The evaporation curves measured for this mixture
are in good agreement with a model
based on an  s-wave Boltzmann equation which neglects inelastic processes.
We have predicted that the scattering interactions are strongly
attractive and widely tunable at low magnetic field.
If the  parameters described above for   deep
wells  can be attained, the system will be  close to the
threshold for superfluidity~\cite{BCS} and  ideal for investigating
frequency shifts and damping in collective
oscillations~\cite{Stringari,Hydrodynamic}. Further,
since s-wave interactions can be
 turned on and off in a few  microseconds, this system is well suited
for studies of many-body quantum dynamics.

This research has been supported
by the Army Research Office and the National Science Foundation.


\begin{references}
\bibitem{Jin1}B. DeMarco, and D. S. Jin, Science {\bf 285}, 1703 (1999).
\bibitem{BCS}H. T. C. Stoof, et al.,
 Phys. Rev. Lett. {\bf 76}, 10 (1996); See also, M. Houbiers
et al., Phys. Rev. A {\bf 56}, 4864 (1997).
\bibitem{Stringari}L. Vichi and S. Stringari, Phys. Rev. A {\bf 60},
4734 (1999).
\bibitem{HighTC}R. Combescot, Phys. Rev Lett. {\bf 83}, 3766 (1999).
\bibitem{Hydrodynamic}G. M. Bruun, and C. W. Clark, Phys. Rev. Lett.
{\bf 83}, 5415 (1999).
\bibitem{Parametric}\mbox{G. M. Bruun and C. W. Clark}, cond-mat/9906392.
\bibitem{Cooperpair}M. Houbiers and H. T. C. Stoof, Phys. Rev. A {\bf
59}, 1556 (1999).
\bibitem{BruunBurnett}G. Bruun, et al., Eur. Phys. J. D {\bf 7}, 433 (1999).
\bibitem{ConMatRev}M. Houbiers and H. T. C. Stoof, cond-mat/9808171.
\bibitem{Elastic}M. Houbiers,  et al., Phys. Rev. A {\bf 57}, R1497 (1998).
\bibitem{s-wave}E. R. I. Abraham, et al., Phys. Rev A {\bf 55}, R3299 (1997).
\bibitem{O'Hara}K. M. O'Hara, et al., Phys. Rev. Lett. {\bf 82}, 4204 (1999).
\bibitem{Verhaar}We are indebted to F. A. van  Abeelen and B. J. Verhaar
who  calculated the  inelastic $|\{3,1\}\rangle\rightarrow
|\{1,2\}\rangle$ dipolar rate and confirmed our calculations of
the  magnetic field dependence of $a_{31}$.
\bibitem{Zoller}P. T\"orm\"a and P. Zoller, Phys. Rev. Lett. {\bf
85}, 487 (2000).
\bibitem{Hansch}S. Friebel, et al., Phys. Rev. A {\bf 57}, R20 (1998).
\bibitem{Magic}{\it Atomic, Molecular, and Optical Physics Handbook},
ed. G. W. Drake, (AIP Press, New York, 1996), p. 176.
\bibitem{Walraven}O. J. Luiten, et al.,
Phys. Rev. A {\bf 53}, 381 (1996).
\bibitem{Chu}S. Chu,  et al., Phys. Rev. Lett. {\bf 55}, 48 (1985).
\bibitem{Adams}C. S. Adams, et al.,  Phys. Rev. Lett. {\bf 74}, 3577 (1995).

\end{references}
\end{document}